\documentclass[10pt]{article}
\usepackage{latexsym}
\usepackage{graphicx}
\usepackage{times}

\begin{document}
\title{Realistic solution to the tunneling time problem}
\author{Wang Guowen\\College of Physics, Peking University, Beijing, China}
\date{June 24, 2007}
\maketitle

\begin{abstract}
There remains the old question of how long a quantum particle
takes to tunnel through a potential barrier higher than its
incident kinetic energy. In this article a solution of the
question is proposed on the basis of a realistic explanation of
quantum mechanics. The explanation implies that the tunneling
particle has a certain chance to borrow enough energy from
self-interference to high-jump over the barrier. The
root-mean-square velocity and the effective tunneling time of an
electron tunneling through a rectangular barrier are numerically
calculated. No superluminal effect (Hartman effect) is found for
the tunneling electron. Heisenberg's energy-time uncertainty
relation for the tunneling effect is verified by calculating an
introduced coefficient representing uncertainty. The present
author argues that phase time, dwell time and
B\"{u}tticker-Landauer time are not appropriate expressions for
the actual transit time in a tunneling process. A quantum
high-jumping model is presented to resolve the paradox that
kinetic energy of the tunneling particle is negative and its
momentum is imaginary.
\end{abstract}

\section{Introduction}
The quantum tunneling through a potential barrier is one of
paradigms of quantum self-interference, which involves
interpretation and application of quantum mechanics. However,
there remains the old question of how long a quantum particle
takes to tunnel through a potential barrier higher than its
kinetic energy. In his 1932 paper [1] L. A. MacColl
concluded:$``$It is found that the transmitted packet appears at
point $x=a$ at about the time at which the incident packet reaches
the point $x=0$, so that there is no appreciable delay in the
transmission of the packet through the barrier.$"$ After three
decades, T. E. Hartman stated: $``$For thicker barriers the peak
of the transmitted packet is shifted, relative to the incident
packet, to higher energy values. The transmission time becomes
independent of barrier thickness and small compared to the
$``$equal time$"$.$"$[2] These statements imply unbounded
tunneling velocities since the thickness may be infinite. Hitherto
there exist various definitions of the tunneling time in
literature [3-5]. Among those, Wigner's phase time [6] and Smith's
dwell time [7] are paid more attention due to their results
showing superluminality.

Like Young's interference, quantum tunneling is a kind of
self-interference behavior. The origin of the behavior is
discussed in detail in Ref.[8]. Briefly, a free quantum particle
can be described in terms of a non-spreading wave packet
consisting of Fourier components in which there is one component,
called characteristic component exclusively related to energy
$E=\hbar \omega$ and momentum $P=\hbar K$ of the particle, that is
exactly an ordinary wave function. The frequencies $\omega_i$ and
wave vectors $k_i$ of the other components as hidden waves have
nothing to do with the Planck constant, so that the wave packet of
this kind never spreads. This non-spreading wave packet on a lower
level is a kind of primary wave packet completely different from
the de Broglie wave packet as secondary wave packet consisting of
components related to different energies and different momenta.
The part outside the peak of the primary wave packet plays a
dramatic role in its self-interference. According to this
realistic view for quantum behavior, we use a monochromatic plane
wave to describe incompletely the incident particle with kinetic
energy $E=mv^2/2$, propose a new solution to the tunneling time
problem and present a high-jumping model of quantum tunneling to
resolve the paradox that kinetic energy of the tunneling particle
is negative and its momentum is imaginary in the classically
forbidden potential barrier.

\section{Tunneling velocity and time of an electron passing through
a rectangular potential barrier}
We consider a non-relativistic electron with kinetic energy
$E=mv^2/2$ tunneling through a one-dimensional rectangular
potential barrier of thickness \emph{d} and height $V_0$ higher
than the incident kinetic energy. We calculate the momentum
distribution of the electron, its root-mean-square (rms) velocity
and effective tunneling time in the barrier region. According to
quantum mechanics, its energy eigenfunction can be split into
three parts:
\begin{equation}
\label{1} \psi_1(x)=e^{ikx}+Re^{-ikx}, \ \mbox{ } x<0, \ \mbox{ }
k=\frac{\sqrt{2mE}}{\hbar}
\end{equation}
\begin{equation}
\label{2} \psi_2(x)=Ae^{\kappa x}+Be^{-\kappa x},\ \mbox{ }0 \leq
x\leq d, \ \mbox{ } \kappa=\frac{\sqrt{2m(V_0-E)}}{\hbar}
\end{equation}
\begin{equation}
\label{3} \psi_3(x)=Se^{ikx},\ \mbox{ } x>d
\end{equation}
in which
\begin{equation}
\label{4} S=\frac{-2i(k/\kappa) e^{-ikd}}{[1-(k/\kappa
)^2]\textrm{sinh}(\kappa d)-2i(k/\kappa)\textrm{cosh}(\kappa d)}
\end{equation}
\begin{equation}
\label{5} A=\frac{Se^{kd}}{2}(1+\frac{ik}{\kappa })e^{-\kappa d}
\end{equation}
\begin{equation}
\label{6} B=\frac{Se^{kd}}{2}(1-\frac{ik}{\kappa })e^{\kappa d}
\end{equation}
\begin{equation}
\label{7} R=(Se^{ikd-\kappa d}-1)(1+ik/\kappa)^2/[1+(k/\kappa)^2]
\end{equation}
The probability flux density of the incident electron is
\begin{equation}
\label{8}
J_{\textrm{in}}=-\frac{i\hbar}{2m}(\psi_{\textrm{in}}^*\frac{\partial\psi_{\textrm{in}}}{\partial
x}-\psi_{\textrm{in}}\frac{\partial\psi_{\textrm{in}}^*}{\partial
x}), ~\psi_{\textrm{in}}=e^{ikx}
\end{equation}

The momentum ($K\hbar$) distribution of the electron in the
barrier region can be calculated by using the following Fourier
~transform [9]:
\begin{equation}\label{9}
\psi(K)=\frac{1}{\sqrt{2\pi}}\int_0^d e^{-iKx}\psi_2(x)\textrm{d}x
\end{equation}
The normalized distribution probability is
\begin{equation}\label{10}
P(K)=\frac{|\psi(K)|^2}{\int_{-K'}^{K'} |\psi(K)|^2\textrm{d}K}
\end{equation}
The root of mean square of \emph{K} is
\begin{equation}\label{11}
K_{\textrm{\scriptsize{rms}}}=\sqrt{\int_{-K'}^{K'}
K^2P(K)\textrm{d}K}
\end{equation}
In this work, we define
$v_{\textrm{\scriptsize{rms}}}=K_{\textrm{\scriptsize{rms}}}\hbar/m$
as the tunneling velocity and define
$t_{\textrm{\scriptsize{eff}}}=d/v_{\textrm{\scriptsize{rms}}}$ as
the effective tunneling time.

As example, we consider the case where $V_0$=10 eV (1
eV=1.6022$\times 10^{-19}$ J), \emph{d}=0.1--1 nm and
\emph{E}=0.1--9.9 eV, use the physical constants: electron mass
\emph{m}=9.1095$\times 10^{-31}$ Kg and Planck constant
$\hbar$=1.055$\times 10^{-34} \textrm{J}\cdot \textrm{s}$ and take
appropriate $K'=7.5\times 10^{10} $/m in calculation. The
calculated results are shown in Fig.1 and Fig.2.
\begin{figure}[htbp]
\centerline{\includegraphics[width=5in,height=2in]{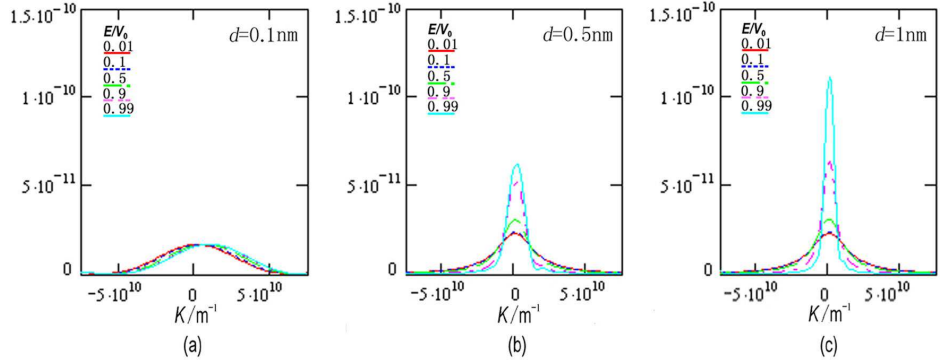}}
\label{Fig.1} \caption{Illustrating the dependence of width of the
momentum ($K\hbar$) distribution on $E/V_0$: (a) For the case of
barrier thickness 0.1 nm the width of the momentum distribution
nearly independent of $E/V_0$. (b) For the case of barrier
thickness 0.5 nm the width of the momentum distribution decreasing
with increasing of $E/V_0$. (c) For the case of barrier thickness
1 nm the width of the momentum distribution rapidly decreasing
with increasing of $E/V_0$.}
\end{figure}
Fig.1(a) shows the width of the momentum distribution nearly
independent of $E/V_0$ for the case of barrier thickness 0.1 nm.
Fig.1(b) shows the width of the momentum distribution decreasing
with increasing of $E/V_0$ for the case of barrier thickness 0.5
nm. Fig.1(c) shows the width of the momentum distribution rapidly
decreasing with increasing of $E/V_0$ for the case of barrier
thickness 1 nm.
\begin{figure}[htbp]
\centerline{\includegraphics[width=5in,height=2.2in]{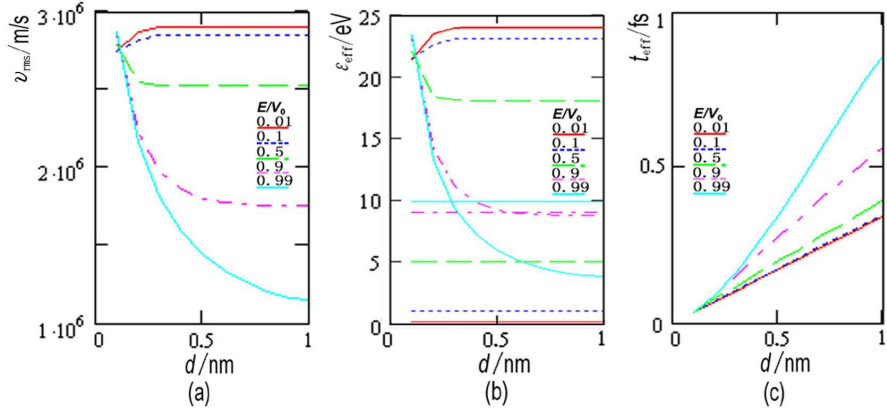}}
\label{Fig.2} \caption{(a) The root-mean-square velocity
($v_{\textrm{\scriptsize{rms}}}$) dependent on the barrier
thickness and $E/V_0$. (b) The effective tunneling kinetic energy
dependent on the barrier thickness and $E/V_0$; the horizontal
lines represent the incident kinetic energies. (c) The effective
tunneling time dependent on the barrier thickness and $E/V_0$.}
\end{figure}
Fig.2(a) shows the root-mean-square velocity
($v_{\textrm{\scriptsize{rms}}}$) dependent on the barrier
thickness and $E/V_0$. Fig.2(b) shows the effective tunneling
kinetic energy ($\varepsilon_{\textrm{\scriptsize{eff}}}$)
dependent on the barrier thickness and $E/V_0$. Fig.2(c) shows the
effective tunneling time ($t_{\textrm{\scriptsize{eff}}}$)
dependent on the barrier thickness and $E/V_0$.

The present definition of the tunneling time is completely
different from the existing typical definitions, such as phase
time [6,10], dwell time [7,10] and B\"{u}tticker-Landauer time
[11]. According to the phase time definition, the tunneling time
is supposed to be:
\begin{equation}\label{12}
t_{\textrm{\scriptsize{ph}}}=\frac{d}{\sqrt{2E/m}}+\hbar
\frac{\textrm{d}(\textrm{arg}S)}{\textrm{d}E}
\end{equation}
Using Eq.4 and this equation for numerical calculation, we obtain
the result as shown in Fig.3(a). Besides, according to the dwell
time definition, the tunneling time is supposed to be:
\begin{equation}\label{13}
t_{\textrm{\scriptsize{dw}}}=\frac{\int_0^d
|\psi_2(x)|^2\textrm{d}x}{|J_{\textrm{in}}|}
\end{equation}
Using Eq.2, Eq.8 and this equation for numerical calculation, we
get the result as shown in Fig.3(b). These results are verified by
using the following analytical expressions [10]:
\begin{equation}\label{14}
t_{\textrm{\scriptsize{ph}}}=\frac{m}{\hbar k\kappa D}[2\kappa
dk^2(\kappa^2-k^2)+(\frac{2mV_0}{\hbar^2})^2\textrm{sinh}(2\kappa
d)],~D=4\kappa^2k^2+
(\frac{2mV_0}{\hbar^2})^2\textrm{sinh}^2(\kappa d)
\end{equation}
\begin{equation}\label{15}
t_{\textrm{\scriptsize{dw}}}=\frac{mk}{\hbar \kappa D}[2\kappa
d(\kappa^2-k^2)+\frac{2mV_0}{\hbar^2}\textrm{sinh}(2\kappa
d)]
\end{equation}
In Fig.3(a) and Fig.3(b) we see that the tunneling times are
independent of the thickness of the thick barrier. This implies
that the tunneling velocity may become even larger than the light
velocity in vacuum since the barrier thickness may be arbitrary
large. This kind of superluminal effect is referred to as Hartman
effect. However, from Fig.2(a) we see no Hartman effect for the
electron. Besides, Fig.3(c) shows B\"{u}tticker-Landauer time
$t_{\textrm{\tiny{BL}}}=md/(\hbar\kappa)$ for an opaque barrier
[11]. Clearly, the phase time, dwell time and
B\"{u}tticker-Landauer time are contradictory to each other. The
present author argues that all these tunneling times are not
appropriate expressions for the actual transit time in a tunneling
process.

\begin{figure}[htbp]
\centerline{\includegraphics[width=5in,height=2.5in]{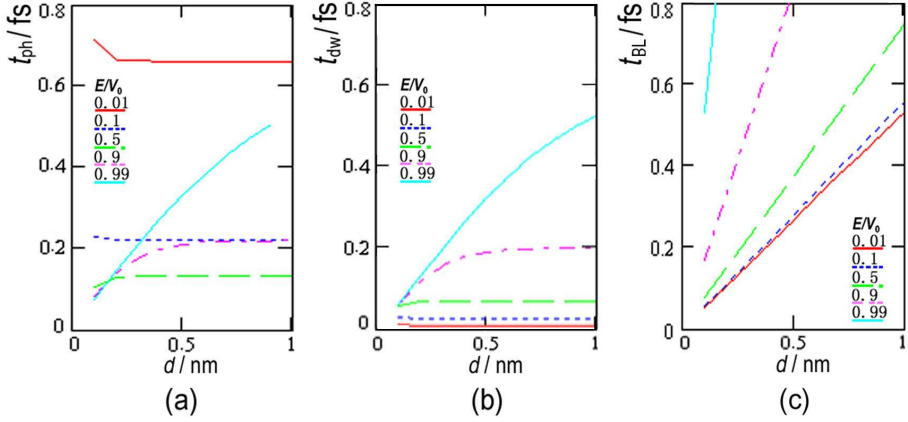}}
\label{Fig.3} \caption{Graphic comparison of three typical
tunneling times: (a) phase time , (b) dwell time, (c)
B\"{u}tticker-Landauer time.}
\end{figure}

\section{Time spent on penetration depth and Heisenberg's energy-time
uncertainty relation} We will shows that Heisenberg's energy-time
uncertainty relation is related to the time spent on penetration
depth into the barrier which can be obtained from the relative
probability density
\begin{equation}
\label{16} D(x)=\frac{|\psi_2(x)|^2}{|\psi_2(0)|^2}, ~0\leq x\leq
d
\end{equation}
when its value reduces to $e^{-2}=0.135$ unless the barrier
thickness is so small that it never reduces to this value in the
barrier region in the case of 0.1 nm thickness as shown in
Fig.4(a). For the cases $E/V_0$=0.01--0.99, the relative
probability densities of the tunneling electron are shown in
Fig.5(a)-(d). The wave function decays approximately exponentially
as a function of \emph{x} in the barrier region. The effective
time spent on the penetration depth \emph{s} is supposed to be
\begin{equation}
\label{17}
\tau_{\textrm{\scriptsize{eff}}}={s}/v_{\textrm{\scriptsize{rms}}}
\end{equation}
Table 1 and Fig.5(a) show the penetration depths. Fig.5(b) shows
the effective times ($\tau_{\textrm{\scriptsize{eff}}}$) spent on
the penetration depths. The curves in Fig.5(c) show the introduced
coefficient ($\xi$) in Heisenberg's energy-time uncertainty
relation, defined by the relation
\begin{equation}
\label{18} \Delta E \Delta t=\xi\hbar/2, ~\Delta
E=\varepsilon_{\textrm{\scriptsize{eff}}}, ~\Delta
t=\tau_{\textrm{\scriptsize{eff}}}
\end{equation}
We see that for $E/V_0$=0.01--0.99 Heisenberg's energy-time
uncertainty relation is satisfied very well since $1.5<\xi\leq 5$
.\\

\begin{figure}[htbp]
\centerline{\includegraphics[width=5in,height=2.5in]{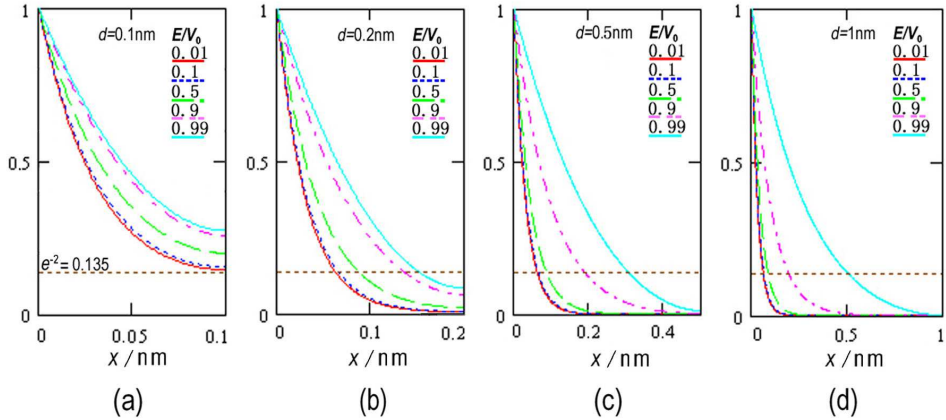}}
\label{Fig.4} \caption{Relative probability densities of the
tunneling electron in the barrier region for the cases: (a)
barrier thickness 0.1 nm, (b) barrier thickness 0.2 nm, (c)
barrier thickness 0.5 nm, (d) barrier thickness 1 nm. The
horizontal line represents the value of $e^{-2}=0.135$.}
\end{figure}

\begin{figure}[htbp]
\centerline{\includegraphics[width=5in,height=2.2in]{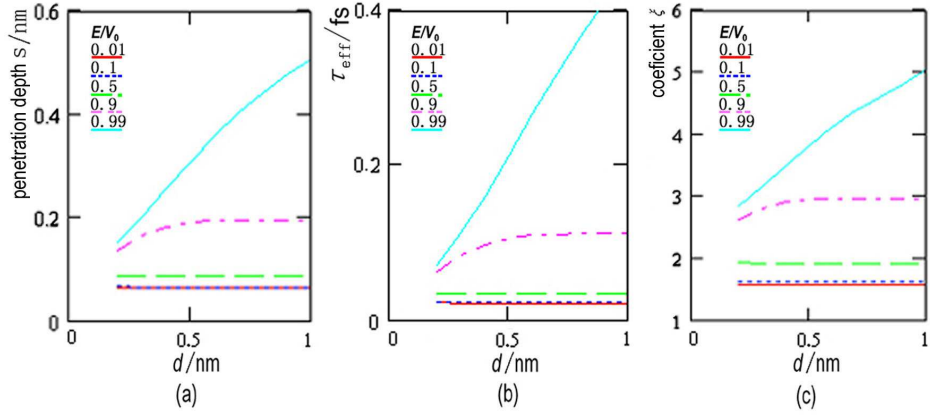}}
\label{Fig.5} \caption{(a) The penetration depth of the electron
in the barrier region. (b) The effective time spent on the
penetration depth. (c) The coefficient in Heisenber's energy-time
uncertainty relation for verifying the relation.}
\end{figure}

\begin{tabular}{|l@{}||c@{}|c@{}|c@{}|c@{}|c@{}|c@{}|c@{}|c@{}|c@{}|}\hline
                    & \multicolumn{9}{|c|}{ barrier thickness \emph{d} (nm) and penetration depth
                     \emph{s} (nm)} \\
\hline    $E/V_0$ & \emph{d}=0.2  & \emph{d}=0.3  & \emph{d}=0.4
& \emph{d}=0.5  & \emph{d}=0.6
& \emph{d}=0.7  & \emph{d}=0.8  & \emph{d}=0.9  & \emph{d}=1.0  \\
\hline\hline 0.01 & 0.0627 & 0.0621 & 0.0621 & 0.0621 & 0.0621 & 0.0621 & 0.0621 & 0.0621 & 0.0621 \\
\hline       0.1  & 0.0658 & 0.0651 & 0.0651 & 0.0651 & 0.0651 & 0.0651 & 0.0651 & 0.0651 & 0.0651 \\
\hline       0.5  & 0.0876 & 0.0874 & 0.0874 & 0.0874 & 0.0874 & 0.0874 & 0.0874 & 0.0874 & 0.0874 \\
\hline       0.9  & 0.1357 & 0.1632 & 0.1811 & 0.1897 & 0.1933 & 0.1947 & 0.1952 & 0.1953 & 0.1954 \\
\hline       0.99 & 0.1521 & 0.2012 & 0.2547 & 0.3069 & 0.3560 & 0.4011 & 0.4413 & 0.4767 & 0.5065 \\
\hline
\end{tabular}\\
Table 1: Penetration depths of the tunneling electron in the
barrier regions

\section{High-jumping model of quantum tunneling}
\begin{figure}[htbp]
\centerline{\includegraphics[width=3.4in,height=3.4in]{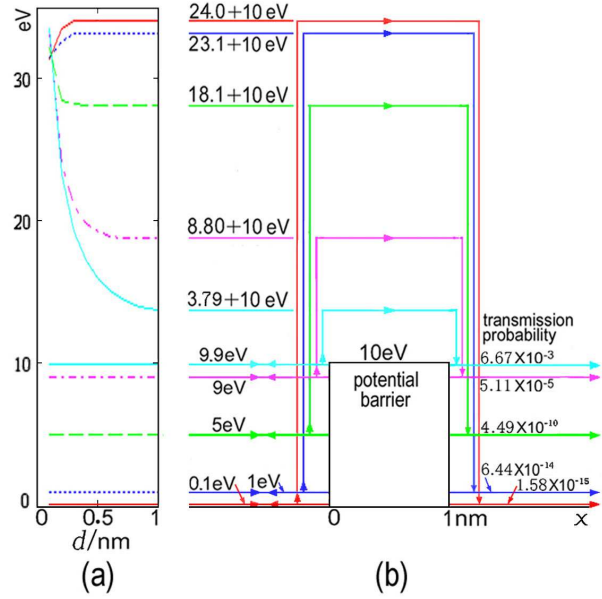}}
\label{Fig.6} \caption{(a) The incident kinetic energy (horizontal
line, 0.1--9.9 eV) and the effective kinetic energy plus the
potential energy of the electron in the barrier region. (b)
Illustration of quantum high-jumping model for the electron
tunneling through the barrier of thickness 1 nm.}
\end{figure}

The quantum tunneling effect can be realistically explained by a
high-jumping model for a quantum particle which has a certain
chance to borrow enough energy from self-interference [8,9] to
jump over a potential barrier as shown in Fig.6(b). We see that
the less the incident kinetic energy of the electron is the higher
it jumps until reaching a limit, for example, limit 34.102 eV for
\emph{d}=1 nm. In this high-jumping model, the kinetic energy is
positive and the momentum is real in the classically forbidden
potential barrier. On the contrary, it is widely accepted in
literature that the former is negative and the latter is imaginary
in the barrier. Realistically, the kinetic energy is never
negative since both mass and velocity squared are always positive.
The quantum high-jumping model demonstrates that the non-peaked
part of the primary non-spreading wave packet describing a quantum
particle plays a dramatic role in its self-interference.

\section{Conclusion}
There remains the old question of how long a quantum particle
takes to tunnel through a potential barrier higher than its
incident kinetic energy. In this article a solution of the
question is proposed on the basis of a realistic explanation of
quantum mechanics. The explanation implies that the tunneling
particle has a certain chance to borrow enough energy from
self-interference to high-jump over the barrier. The
root-mean-square velocity and the effective tunneling time of an
electron tunneling through a rectangular barrier are numerically
calculated. No superluminal effect (Hartman effect) is found for
the tunneling electron. Heisenberg's energy-time uncertainty
relation for the tunneling effect is verified by calculating an
introduced coefficient representing uncertainty. The present
author argues that phase time, dwell time and
B\"{u}tticker-Landauer time are not appropriate expressions for
the actual transit time in a tunneling process. A quantum
high-jumping model is presented to resolve the paradox that
kinetic energy of the tunneling particle is negative and its
momentum is imaginary.

\end{document}